\documentclass[aps,showpacs,preprint,preprintnumbers,amsmath,amssymb,nofootinbib,contrib]{revtex4}
\usepackage{epsfig}
\usepackage{graphicx}
\usepackage{dcolumn}
\usepackage{bm}

\def\beq{\begin{equation}}
\def\eeq{\end{equation}}

\newcommand{\bea}{\begin{eqnarray}}
\newcommand{\eea}{\end{eqnarray}}

\def\eeqn{\end{equation}}
\newcommand\iden{\leavevmode\hbox{\small1\normalsize\kern-.33em1}}

\let\jnfont=\rm
\def\NPB#1,{{\jnfont Nucl.\ Phys.\ B }{\bf #1},}
\def\PLB#1,{{\jnfont Phys.\ Lett.\ B }{\bf #1},}
\def\EPJC#1,{{\jnfont Eur.\ Phys.\ Jour.\ C }{\bf #1},}
\def\PRD#1,{{\jnfont Phys.\ Rev.\ D }{\bf #1},}
\def\PRL#1,{{\jnfont Phys.\ Rev.\ Lett.\ }{\bf #1},}
\def\MPLA#1,{{\jnfont Mod.\ Phys.\ Lett.\ A }{\bf #1},}
\def\JPG#1,{{\jnfont J.\ Phys.\ G }{\bf #1},}
\def\CTP#1,{{\jnfont Commun.\ Theor.\ Phys.\ }{\bf #1},}
\def\JHEP#1,{{\jnfont JHEP \ }{\bf #1},}
\def\NPPS#1,{{\jnfont Nucl.\ Phys.\ Proc.\ Suppl.\ }{\bf #1},}

\newcommand{\etal}{{\it et al.}}

\begin{document}


\title{Productions of heavy charged leptons via gluon fusion at LHC \\
: A revisit}

\author{Chun Liu and Shuo Yang}
\affiliation{Key Laboratory of Frontiers in Theoretical Physics,
Institute of Theoretical Physics, Chinese Academy of Sciences,
P.O. Box 2735, Beijing 100190, China }
\email{liuc@mail.itp.ac.cn, shuoyang@itp.ac.cn}

\begin{abstract}
Heavy charged lepton productions via gluon fusion at the LHC are
revisited.  Full loop calculations are adopted with an updated
parton distribution function and electroweak data.  Including
contribution from new generation quarks in the loop, pair production
of the sequential heavy lepton via gluon fusion at the LHC dominates
over that via the Drell-Yan mechanism in some heavy lepton mass
range. Exotic lepton single production of vectorlike lepton extended
models is also calculated.  In the later case, the gluon fusion
mechanism via the Higgs exchange is emphasized.  Our numerical
results for both pair and single production of heavy leptons are
smaller than previous studies especially for a large heavy lepton
mass as a result of full loop calculation and due to the mixing
angles.
\end{abstract}

\pacs{14.60.Hi, 12.60.-i, 13.85.Qk }

\keywords{heavy lepton, LHC}
\maketitle

\section{Introduction}

The CERN Large Hadron Collider (LHC) is the highest energy physics
experiment of our time. In addition to the Higgs particle which is
the last necessity of the Standard Model (SM), its main goal is
searching for the physics beyond the SM. Imaginable new physics
discoveries at the LHC can be new fermions, new gauge bosons, extra
Higgs and so on.  Among these possibilities, we study new charged
leptons.  Although new lepton observation maybe challenging at the LHC,
once they are produced, their decay signals are easy to be identified.

The new charged leptons are introduced in many new physics models
such as grand unification theories, mirror fermions, supersymmetry
and little higgs. In some models, new fermions play an important
role in electroweak symmetry breaking or CP violation, and their
characters may be different from the presently known fermions.
Discovery of such new fermions would revolutionize our understanding
of electroweak symmetry breaking and some other basic problems.

At hadron colliders, the Drell-Yan process \cite{dy} and the gluon
fusion process \cite{Scottgg} are expected to be the main mechanisms
of heavy charged lepton production.  In the extreme case that the
new leptons are vectorlike and have no Yukawa interactions, the
Drell-Yan mechanism is dominantly responsible for new lepton
production.  On the other hand, if the new leptons are chiral with
large Yukawa couplings, their production through Higgs mediated
processes can be significant, and the virtual Higgs is produced via
gluon fusion.  Due to the large rate of gluons at the LHC, as well
as the new contributions from new quarks, the gluon fusion
production can dominate over the Drell-Yan mechanism for new chiral
charged leptons in some parameter region.  This was also studied in
refs. \cite{Framgg,4,singlelepton,vectorpheno,pairNLO,review}.  In
an effort of understanding the Higgs, an extra vectorlike generation
of matter is introduced within the framework of supersymmetry
\cite{Liu}. What is new in the lepton sector of that model after
supersymmetry breaking is a vectorlike SU(2)$_L$ singlet lepton with
a mass of $\sim\mathcal{O}(400)$ GeV.  We are interested in looking
at its production at the LHC.  There are other vectorlike extensions
to the minimal supersymmetric SM \cite{vectorlike}. Generally, heavy
leptons, even if they are vectorlike, have Yukawa interactions which
may enhance their production rates.  It is the gluon fusion
mechanism which is the focus of this study.

In view of the current knowledge about parton distribution function,
relevant electroweak data and the top quark mass, the old results of
heavy lepton production via gluon fusion should be updated.  Furthermore
previous studies on the gluon fusion mechanism for lepton production
took tree level approximation.  We will update previous studies about
heavy charged lepton production via the gluon fusion mechanism in
complete loop calculation.  It is found that the tree level
approximation should be carefully used in heavy lepton production from
the gluon fusion mechanism and it is only valid in some limits.

This paper is organized as follows. In Sec. II, simple heavy fermion
scenarios and their phenomenological constraints are described.  In
Sec. III, pair production of sequential charged fermions are
calculated. Sec. IV discusses vectorlike fermion extension of the
SM, and single production of the exotic fermion in this scenario.
Finally, we make a discussion and give our conclusions in Sec. V.


\section{The new leptons}

New fermions appear in various new physics models.  They can be
classified to be chiral or vectorlike. In this section, we will
start with a description of these two scenarios of new fermions and
then discuss the phenomenological bounds.

One can make a replica of a SM family to get the simplest fourth
generation which is the so-called sequential fermions
\cite{newreport}.\footnote{The invisible width of Z boson and the
direct search limit require that the fourth neutrino must be heavy.
So a sequential fourth generation should also include a single
right-handed neutrino $\nu_{4R}$. We will not discuss the collider
phenomenology of the new neutrino in this paper.}  The sequential
new leptons $L_4$, $E^c_4$ fall into the representations $(2,-1)$,
$(1,2)$ under $SU(2)_L\otimes U(1)_Y$, respectively.

The new leptons can also be vectorlike, where the left and right
components transform the same under $SU(2)_L\otimes U(1)_Y$.  Both
vectorlike doublet leptons and vectorlike singlet leptons are simple
examples \cite{review}. The quantum numbers for vectorlike singlet
lepton pair $e_{4}$, $e^c_{4}$ are $(1,-2)$, $(1,2)$, respectively.
And those for vectorlike doublet leptons $L_{4}$, $L_{4}^c$ are
$(2,-1)$, $(2,1)$.

Now let us consider phenomenological constraints to these scenarios.
The direct experimental search of new leptons at LEPII requires that
the new charged leptons should be heavier than 102 GeV and fourth
neutrino heavier than 101 GeV \cite{lep}.  The results for the pure
Dirac neutrino and for the neutrino with a Majorana mass are
slightly different.  As for new quarks, the strongest bound on $u_4$
is $m_{u_4}>256$ GeV \cite{CDFquark}, which comes from CDF by
searching for $u_4\bar{u}_4$ with $u_4$($\bar{u}_4$) decay to
$W^+$($W^-$) boson and an ordinary quark.  Assuming the branching
ratio $BR(d_4 \rightarrow bZ)=1$, CDF obtains the bound
$m_{d_{4}}>268$ GeV \cite{CDFbprime}.  Additionally, the constraints
from the Z width require that new fermion masses are larger than
$M_Z/2$ which are weaker than those from the direct search.

For sequential fermions, the most stringent constraints are from
"oblique parameters" $S$, $T$ and $U$ \cite{PDG}. These constraints
can be relaxed by allowing $T$ to vary or fourth generation masses
are not degenerate \cite{constraints,4thandH}.  Recently, ref.
\cite{4thandH} has identified a region for new sequential fermions
which agrees with all experimental constraints and has minimal
contributions to oblique parameters.  In this paper, we will assume
a similar parameter as that in \cite{4thandH}.  Flavor physics also
gives constraints on the fourth generation.  Mixing parameters
between the extra fermions and the ordinary three generations are
subject to processes such as $\mu \rightarrow e \gamma$ decay and
$D^0-\bar{D}^0$ mixing.  These constraints \cite{PDG} are strong on
the mixing between the first or second generation and the fourth
generation, which suggest that mixings need to be smaller than 0.01.
For the mixing between the third generation and the fourth
generation, the flavor constraints are not very strong.

As for vectorlike extensions, the most important consequence is the
flavor changing neutral current (FCNC).  Because of introducing
vectorlike fermions, there is no GIM mechanism to suppress the FCNC
related to these fermions.  Furthermore, there is a resultant effect
on flavor diagonal neutral currents \cite{bargersinglet}.  The decay
width of Z boson forces this effect to be small.  This constraints
the mixing angles strongly.  Vectorlike fermions do not contribute
to "oblique parameters" in the leading order, and thus these
parameters do not constrain their masses.

\section{Pair production of charged sequential heavy lepton}

Within the framework of the fourth chiral generation, pair
production is the main interest about heavy charged leptons.  In
addition to that via the Drell-Yan process, the heavy leptons can
also be produced via the gluon fusion mechanism induced by fermion
loops as shown in Fig. 1.  And this mechanism could dominate over
the Drell-Yan mechanism in some parameter space due to the large
rate of gluons at the LHC \cite{Scottgg,Framgg}.  There is no photon
exchange diagrams, and only the Higgs and the Z boson with axial
vector coupling contributes to this gluon fusion due to Furry's
theorem. As the ggH vertex in Higgs exchange diagram is a symmetric
tensor while ggZ vertex in Z exchange diagram is antisymmetric,
there is no interference between these two contributions. In this
section, we will study the pair production of the sequential lepton
via the Higgs exchange diagram and the Z exchange diagram
separately.

\begin{figure}[htbp]
 \includegraphics[width=3in]{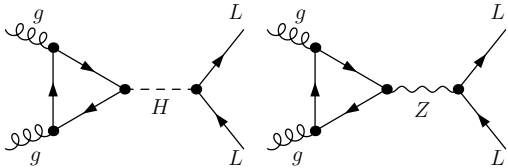}
\caption{Feynman diagrams for heavy lepton pair production via gluon
fusion. The gluon crossing diagrams are not shown.}
\end{figure}

\begin{figure}[htbp]
 \includegraphics[width=3in]{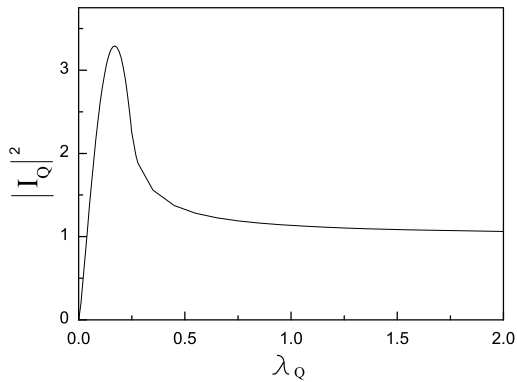}
\caption{$|I_{Q}|^2$ as a function of $\lambda_Q$ where $I_Q$ is the
ggH effective loop function}
\end{figure}

The Higgs exchange diagram for heavy lepton production in Fig.1 is
related to Higgs production in the gluon-gluon fusion mechanism. The
ggH effective Lagrangian can be presented as ${\mathcal
L}=\displaystyle\frac{\alpha_s}{12\pi} G^{\mu\nu}G_{\mu\nu}H I_{H}$
where $I_H$ is following loop function \cite{IggH}, \beq
 I_H=\sum_Q I_Q, \ \ \ I_Q=3\int^1_0 dx\int^{1-x}_0 dy
 \frac{1-4xy}{1-xy/(m_Q^2/\hat{s})-i\epsilon}\,.
\eeq
Replacing the C. M. energy of subprocess $\sqrt{\hat{s}}$ by $m_H$,
one can get the loop function for Higgs production.  In general, the
loop function $I_H$ is complex and evaluation of
the integral gives $I_{Q}$ in terms of $\lambda_Q=m_{Q}^{2}/\hat{s}$,
\beq
I_{Q}=[2\lambda_Q+\lambda_Q(4\lambda_Q-1)f(\lambda_Q)]
\eeq
where
\begin{eqnarray}
 f(\lambda)  =  \left\{
\begin{array}{ll} \displaystyle
    -2[\sin^{-1}\frac{1}{2\sqrt{\lambda}}]^2
    & \qquad {\rm for\ \lambda > \frac{1}{4}}\,, \\[3mm]
\displaystyle
\frac{1}{2}
[\ln(\frac{1+\sqrt{1-4\lambda}}{1-\sqrt{1-4\lambda}})-i
 \pi]^2
    & \qquad {\rm for\ \lambda < \frac{1}{4}}\,. \\[3mm]
\end{array}
\right.
\end{eqnarray}
For convenience in discussion, we show the curve of $|I_Q|^2$ as a
function of $\lambda_Q$ in Fig.2. \footnote{There is a similar
diagram and a detailed discussion of $I_H$ in \cite{Colliderbook}}
When $m_{Q}$ is much heavier than $\sqrt{\hat{s}}$, i.e.
$\lambda_Q\gg 1$, $I_{Q}$ reaches 1 which is just the so-called the
heavy top quark limit for light Higgs production via the gluon
fusion mechanism. In the small $m_{Q}$ limit $\lambda_{Q}\ll 1$,
$I_{Q} \rightarrow 0$. There is also a peak for $|I_Q|^2$ at that
$|I_Q|^2\simeq 3.3$ for $\lambda_Q$ being 0.17.  For the process
gluon-gluon fusion to a light Higgs where $\sqrt{\hat{s}}=m_H$, it
is correct to take limit $\lambda_Q=m_{top}^2/m_H^2\gg 1$ for the
top quark and $\lambda_Q=m_q^2/m_H^2\ll 1$ for light quarks.
However, when it turns to heavy lepton pair production, the
subprocess C.M. energy $\sqrt{\hat{s}}$ varies from $4m_L^2$ where
$m_L$ is the mass of the heavy lepton to several TeV, and thus the
$\lambda_Q$ is not fixed. In ref. \cite{Scottgg}, it was assumed
that $I_Q$ receives a value of unity from every quark with $m_Q>
m_L$, which is a rough approximation. However, in some later studies
$I_Q$ was taken to be unity irrespective of the relation between
$m_Q$ and $m_L$ and the variance of $\sqrt{\hat{s}}$.  That is
unreasonable and would overestimate the cross section for large
$m_L$.  In fact the effective function $I_H$ should be carefully
used for different $\lambda_Q$ and it is better to calculate the
cross section in loop for dilepton production from gluon fusion. We
deduce the interaction vertices of ggH and ggZ and express them in
terms of Passarino-Veltman scalar loop functions \cite{Sfun}. The
cross sections are calculated in completed loop calculation with
LoopTools \cite{LoopTools}. Detailed representations are shown in
appendix. We have used CTEQ6L \cite{cteq6} parton distribution
function with factorization scale $\mu_f=2m_L$. The input parameters
relevant to our computation are $m_t=172.7$ GeV \cite{topmass},
$m_b(m_b)=4.2$ GeV, $m_Z=91.1876$ GeV, $\sin^2\theta_W=0.2315$,
$\alpha_{e}(M_Z) = 1/128.8$ and the two-loop running coupling
constant $\alpha_s(M_Z) = 0.1176$ \cite{PDG}.

\begin{figure}[htbp]
 \includegraphics[width=3.5in]{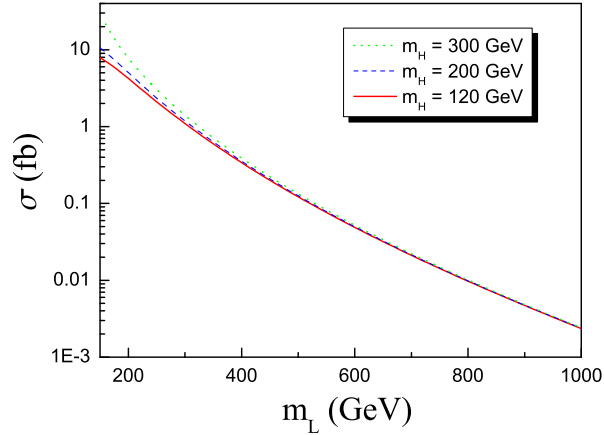}
\caption{Cross section for sequential heavy lepton pair production
from Higgs exchange diagrams with only third generation quarks in the
loops as a function of lepton mass $m_L$ for
$m_H=120$ GeV (solid line), $m_H=200$ GeV (dash line), $m_H$=300
GeV(dot line).}
\end{figure}

\begin{figure}[htbp]
 \includegraphics[width=3.5in]{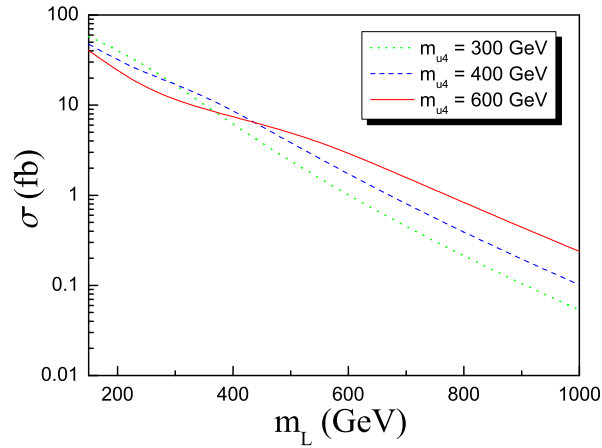}
\caption{Cross section for sequential heavy lepton pair production
from Higgs exchange diagrams as a function of lepton mass $m_L$ for
$m_H=120$ GeV and
$m_{u_4}=600$ GeV (solid line), $m_{u_4}=400$ GeV (dash line),
$m_{u_4}$=300 GeV (dot line).}
\end{figure}

Fig.3 plots the cross section for new sequential lepton pair
production at the LHC with $\sqrt{s}=14$ TeV versus the mass
parameter $m_L$ of the new charged lepton for several choices of the
Higgs mass $m_H$.  Here only contributions from the top quark and
the bottom quark are included. Actually, the bottom quark's
contribution is tiny in heavy lepton pair production as
$m_b/\sqrt{\hat{s}}\rightarrow 0$ for most of $\sqrt{\hat{s}}$,
which is different from the case of Higgs production.  It is found
that the cross section is sensitive to the square of the mass of the
new lepton. And it can be enhanced by a heavy Higgs mass especially
for a light new lepton.  For a typical new lepton mass $m_L=200$ GeV
and the Higgs mass $m_H=300$ GeV ($m_H=120$ GeV), the cross section
is 7.8 fb (4.2 fb).

We also take into account the contributions from new generation
sequential quarks in the loop for $m_H=120$ GeV, which enhances
significantly the cross section in all of the parameter space as
shown in Fig.4. Several typical heavy quark mass parameter values
$m_{u_4}=$ 300 GeV, 400 GeV, 600 GeV and the relation
$m_{d_4}=m_{u_4}-50$ GeV which agree well with current data
\cite{4thandH} are used.  The loop function of ggH interaction does
not monotonously depend on $\sqrt{\hat{s}}$, the effects due to new
quarks are complicated as can be seen in Fig. 4. For fixed $m_L$
with $ m_{u_4}^2/4 m_L^2<0.17 $, the value of
$\lambda_{u_4}=m_{u_4}^2/\hat{s}$ is smaller than 0.17 for all
$\hat{s}$ and the loop function $I_Q$ is in the monotonous region.
The heavy quark effect in this region is more important than that in
the other region. It is found that the contributions from the new
generation quarks are significant.  For $m_{u_4}$=400 GeV and
$m_L=200$ GeV, the cross section is 32 fb.  Even for a larger mass
$m_L=500$ GeV, the cross section is still as large as 3.85 fb.
Unlike the case of light Higgs production via gluon fusion where a
generation of quarks increases the cross section by roughly a factor
of 9 \cite{4thandH, 4thHiggs}, in lepton pair production the
increase is smaller than 9 times in low $m_L$ region but it is much
larger than 9 times in high $m_L$ region.

Now we consider the Z exchange diagram. The ggZ interaction vertex
can be expressed as \cite{IggZ}: \bea F^{\alpha\mu\nu}=\sum_Q
\frac{g_ag_s^2Tr[T^aT^b]}{4\pi^2}
[\varepsilon^{\mu\nu\omega\varphi}p_{\omega}q_{\varphi}k^{\alpha}
F_1(k^2)+(\varepsilon^{\alpha\mu\omega\varphi}q^{\nu}
-\varepsilon^{\alpha\nu\omega\varphi}q^{\mu})p_{\omega}q_{\varphi}
F_2(k^2) \\ \nonumber +(\varepsilon^{\alpha\mu\omega\varphi}p^{\nu}
-\varepsilon^{\alpha\nu\omega\varphi}q^{\mu})p_{\omega}q_{\varphi}
F_3(k^2)+(\varepsilon^{\alpha\mu\nu\omega}(p_{\omega}-q_{\omega})]
F_4(k^2) \label{ggZ} \eea where $g_a$ is the coupling of axial
vector current and $F_i(k^2)$'s ( $i=1-4$ ) are scalar functions,
\bea
 &&F_1=\int^1_0 dx\int^{1-x}_0 dy\ [m_Q^2-k^2xy]^{-1}[(x+y)(1-x-y)+4xy]\\
 -&&F_2=F_3=\int^1_0dx\int^{1-x}_0dy\ [m_Q^2-k^2xy]^{-1}[(x+y)(1-x-y)]\\
 &&F_4=1+\int^1_0dx\int^{1-x}_0dy\ [m_Q^2-k^2xy]^{-1}[-2(m_Q^2-k^2xy)
+1/2 k^2(x+y)(1-x-y)] \label{fterms}\eea where the unity in $F_4$ is
the anomaly term.
\begin{figure}[htbp]
 \includegraphics[width=3.5in]{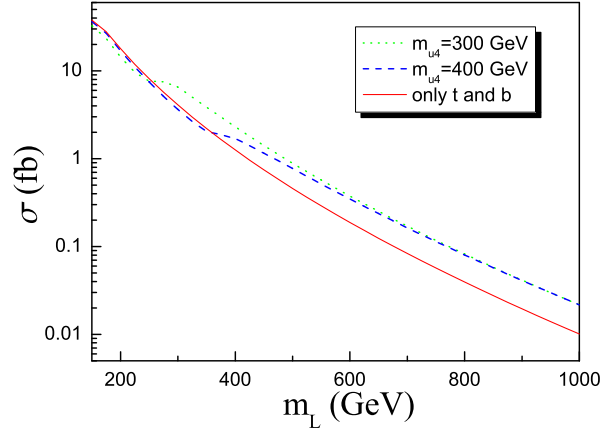}
\caption{Cross section for sequential heavy lepton pair production from
the Z exchange diagrams as a function of lepton mass $m_L$ for
$m_{u_4}=300$ GeV (dot line), $m_{u_4}=400$ GeV (dash line) and
that with the third generation quarks only (solid line).}
\end{figure}

\begin{figure}[htbp]
 \includegraphics[width=3.5in]{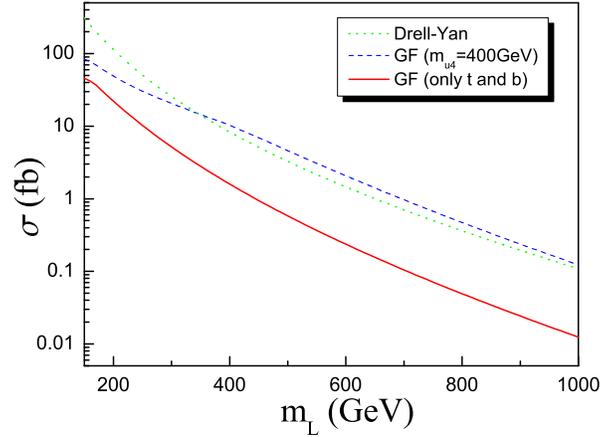}
\caption{Cross section for sequential heavy lepton pair production via
the Drell-Yan mechanism (dot line), the gluon fusion mechanism without
contribution from new quarks (solid line) and with contribution
from new quarks where $m_{u_4}=400$ GeV, $m_{d_4}=m_{u_4}-50$ GeV (dash
line).}
\end{figure}

Because of the different signs of axial vector coupling for up-type
and down-type quarks, the contributions from up-type and down-type
quarks are destructive. For the first two generations, the mass
split between up-type and down-type quarks $\Delta m_Q \sim 0$, so
the total contribution from the first two generations is vanishing.
The cross section with top quark and bottom quark contribution only
is shown in the solid line in Fig. 5.  The cross section with top
quark and bottom quark contribution only is larger than that of the
corresponding Higgs exchange diagrams. For the heavy lepton with a
mass from 250 GeV to 150 GeV, the cross section can reach 8.1 - 38
fb.  We have also considered contribution from the new quarks with
$m_{u_4}=$ 300 GeV or 400 GeV and $m_{d_4}=m_{u_4}-50$ GeV.  The
role of the new quarks is significant for a larger $m_L$. Generally,
a larger split between new generation quarks will result in a higher
production rate in Z exchange diagrams, which was used in previous
studies, but a very large split is conflict with phenomenological
constraints.

The total cross section for heavy lepton pair production via gluon
fusion  $\sigma_{gg}$
 is the sum of the contributions from the Z exchange diagram and
the Higgs exchange diagram.  In Fig. 6, this cross section is
compared with that via the $q\bar{q}\rightarrow \gamma/Z \rightarrow
LL$ Drell-Yan mechanism $\sigma_{DY}$.  If only the third generation
quarks are considered, $\sigma_{gg} < \sigma_{DY}$. By taking into
account the new generation quarks, $\sigma_{gg}$ dominates over
$\sigma_{DY}$ in the large mass region. For instance, assuming
$m_{u_4}=400$ GeV and $m_{d_4}=m_{u_4}-50$ GeV, $\sigma_{gg} >
\sigma_{DY}$ for the heavy lepton mass ranging from 350 GeV to 1000
GeV.  Our numerical results about gluon fusion are smaller than
previous studies\cite{Scottgg,Framgg} especially for the large heavy
lepton mass. This is mainly because we have used full loop
calculation and due to the axial couplings.

The total cross section of heavy lepton pair production is
enhanced significantly which increases the possibility of detecting
the heavy lepton signal.
With a luminosity 100 fb$^{-1}$, including contributions from new
generation quarks, we predict that for the sequential lepton
mass $m_L=250$ GeV, 8100 heavy charged lepton pair events can be
produced at the LHC with $\sqrt s$=14 TeV.  If heavy charged lepton
mass $m_L$ is larger than heavy neutrino mass $m_{\nu_L}$, the main
decay modes of heavy charged lepton are $L \rightarrow \nu_L W^*
\rightarrow \nu_Ll\bar{\nu_l}$ and $L \rightarrow \nu_L W^*
\rightarrow \nu_L q\bar{q}'$. In the other case $m_L<m_N$
\cite{hou}, $L$ will only decay via Cabibbo-suppressed $L
\rightarrow \nu_{\tau}W^*$ with leptonic and hadronic decay of
$W^*$. \footnote{For $m_L<m_N$, the most promising detecting mode of
heavy lepton is $L\bar{\nu}_L \rightarrow
W^-\bar{\nu}_L\bar{\tau}W^- $. \cite{hou}}
 Assuming $m_L>m_{\nu_L}$ and the fourth generation neutrino is
massless, early work \cite{han} argued that the heavy lepton signal
is buried by standard model backgrounds which mainly are single and
pair production of weak bosons at the SSC with $\sqrt{s}=40$ TeV.
However, as discussed in Sec. II, current constraints require that
fourth generation neutrino holds a large mass which results in
different kinematic distributions of the signal as that in ref.
\cite{han}. If considering the large contributions from new quarks
and using some kinematic tricks, it is hopeful to detect the heavy
lepton signal in some lower $m_L$ region at the LHC. Further
detailed studies are needed.

\section{Single production of exotic leptons in vectorlike extended
models}

For vectorlike fermions via the gluon fusion mechanism, both single
production \cite{singlelepton} and pair production
\cite{Framgg,4,vectorpheno} are possible.  Because single production
has a larger rate than pair production, we consider heavy lepton
single production in this work. Both Drell-Yan processes
\cite{Framgg,singlelepton} and gluon fusion processes
\cite{singlelepton} are involved in the single production.  While
ref. \cite{singlelepton} considered the Z boson mediated gluon
fusion process, we also include the Higgs boson mediated gluon
fusion.  This can be important due to relevant large Yukawa
couplings.  In addition, the third generation quarks in the loop are
considered.  Our calculation also uses full loop calculation
together with updated parton distribution function and electroweak
data.

The single heavy lepton production via the gluon fusion processes is
distinguishable from that via the W boson mediated Drell-Yan processes,
besides the charged heavy lepton, the gluon fusion process also produces
an ordinary charged lepton which can be identified experimentally in
principle.  Nevertheless we will compare the gluon fusion results with
the Z boson mediated Drell-Yan results.

For singlet vectorlike extension, a lepton pair $e_{4}$ and
$e^c_{4}$, with quantum numbers $(1,-2)$ and $(1,2)$ under
$SU(2)_L\otimes U(1)_Y$ are introduced.  For convenience, we only
consider mixing of the third generation and the new vectorlike
fermion.  We write down the Lagrangian relevant to lepton masses,
\begin{align}
{\mathcal L} \supset y_{33}L_3 \tau_2 \Phi^* e_3^c+ f e_{4}e_{4}^c+
y_{34}L_3 \tau_2 \Phi^ *e_4^c +\text{h.c.},
\end{align}
where $y_{33}$ and $y_{34}$ denote Yukawa couplings, $\Phi$ is the Higgs
doublet, $L_3=\begin{pmatrix} \nu_3 \\ l_3\end{pmatrix}$ and $e_3^c$ are
the third generation lepton doublet and singlet, respectively.  Note
that there is no $y_{43}$ term in the formula.  After electroweak
symmetry breaking $\langle\Phi\rangle=\displaystyle\frac{1}{\sqrt{2}}
\begin{pmatrix} 0 \\ v \end{pmatrix}$,
\begin{equation}
{\mathcal L} \supset -\left(l_3, e_4\right) {\mathcal M}^l
\left(
\begin{array}{c}
e_3^c \\ e_4^c
\end{array}
\right) \,.
\end{equation}
The charged lepton mass matrix is given as
\begin{equation}
{\mathcal M}^l = \left(
\begin{array}{cc}
m_{33} & m_{34} \\
0      & f      \\[3mm]
\end{array}
\right) \,,
\end{equation}
where $m_{33}=\displaystyle\frac{y_{33}v}{\sqrt{2}}$ and
$m_{34}=\displaystyle\frac{y_{34}v}{\sqrt{2}}$.  This matrix is
diagonalized by two orthogonal matrices,
\beq
\begin{pmatrix} \cos\theta_L & -\sin\theta_L \\
                \sin\theta_L & \cos\theta_L \end{pmatrix}
\begin{pmatrix} m_{33} & m_{34} \\
                0      & f \end{pmatrix}
\begin{pmatrix} \cos\theta_R  & \sin\theta_R \\
                -\sin\theta_R & \cos\theta_R \end{pmatrix}
=\begin{pmatrix} m_{\tau} & 0 \\ 0 & m_{L} \end{pmatrix} \,,
\eeq
the physical $\tau$ lepton and the new heavy lepton $L$ are,
\bea
\tau &=& \cos\theta_L l_3-\sin\theta_L e_4\,, \qquad
         \tau^c=\cos\theta_R e_3^c-\sin\theta_R e_4^c\,, \\ \nonumber
  L  &=& \sin\theta_L l_3+ \cos\theta_L e_4\,, \qquad
          L^c=\sin\theta_R e_3^c+ \cos\theta_R e_4^c\,.
\eea
The corresponding masses and mixing parameters are
\bea
m_{\tau}^2 &=& \frac{1}{2}\left(f^2+m_{33}^2+m_{34}^2
               -\sqrt{(f^2-m_{34}^2- m_{33}^2)^2+4m_{34}^2 f^2}\right)
               \simeq m_{33}^2\,, \\[3mm] \nonumber
m_L^2      &=& \frac{1}{2}\left(f^2+m_{33}^2+m_{34}^2
               +\sqrt{(f^2-m_{34}^2-m_{33}^2)^2+4m_{34}^2f^2}\right)
               \simeq f^2+m_{34}^2\,; \\[3mm] \nonumber
\sin\theta_L &=& \frac{1}{\sqrt{2}}\sqrt{1-\frac{f^2-m_{34}^2-m_{33}^2}
                 {\sqrt{(f^2-m_{34}^2-m_{33}^2)^2+4 m_{34}^2 f^2}}}
                 \simeq \frac{m_{34}}{f}\,, \\[3mm] \nonumber
\sin\theta_R &=& \frac{1}{\sqrt{2}}\sqrt{1-\frac{f^2+m_{34}^2
                 -m_{33}^2}{\sqrt{(f^2+m_{34}^2-m_{33}^2)^2
                 +4m_{34}^2 m_{33} ^2}}}\simeq\frac{m_{33}m_{34}}{f^2}\,.
\eea
Taking $f>m_{34},m_{33}$, we have made an expansion to order of $v/f$
and keep only leading non-vanishing results.

Now let us turn to the doublet vectorlike fermions.  The vectorlike
doublet extension introduces a doublet lepton pair $L_{4}$ and
$L^c_{4}$ with quantum numbers $(2,-1)$ and $(2,1)$ under $SU(2)_L
\otimes U(1)_Y$. The Lagrangian relevant to the mass is:
\begin{align}
{\mathcal L} \supset y_{33}L_3 \tau_2 \Phi^* e_3^c+ f L_{4}L_{4}^c
+y_{43}L_4 \tau_2 \Phi^*e_3^c +\text{h.c.}\,.
\end{align}
As in the case of the vectorlike singlet model, the masses and
mixing parameters are obtained, \bea m_{\tau}^2 &=&
\frac{1}{2}\left(f^2+m_{33}^2+m_{43}^2
               -\sqrt{(f^2+m_{43}^2-m_{33}^2)^2+4m_{43}^2f^2}\right)
               \simeq m_{33}^{2} \,, \\[3mm] \nonumber
m_L^2      &=& \frac{1}{2}\left(f^2+m_{33}^2+m_{43}^2
               +\sqrt{(f^2+m_{43}^2-m_{33}^2)^2+4m_{34}^2f^2}\right)
               \simeq f^2+m_{43}^2 \,; \\[3mm] \nonumber
\sin\theta_L &=& \frac{1}{\sqrt{2}}\sqrt{1-\frac{f^2+m_{43}^2-m_{33}^2}
                 {\sqrt{(f^2+m_{43}^2-m_{33}^2)^2+4m_{43}^2m_{33}^2}}}
                 \simeq \frac{m_{43}m_{33}}{f^2} \,, \\[3mm] \nonumber
\sin\theta_R &=& \frac{1}{\sqrt{2}}\sqrt{1-\frac{f^2-m_{43}^2-m_{33}^2}
                 {\sqrt{(f^2-m_{43}^2-m_{33}^2)^2+4m_{43}^2 f^2}}}
                 \simeq \frac{m_{43}}{f}\,.
\eea

The interaction vertices are obtained after replacing the weak
eigenstates by the physical states.  Higgs-fermion-fermion and
Z-fermion-fermion interactions for physical $\tau $ and $L$ are
listed in Table I.  The feynman rules for $ZLL$ $ZL\tau$ agree
with that given in ref. \cite{feynmanvector}.  The $L$ related
tree level FCNC is explicitly seen.

\begin{table}[h]
\begin{tabular}{|c|c|c|}
\hline \hline & Vectorlike singlet model& Vectorlike doublet model
\\ \hline
$H \bar{\tau} \tau$ & $\displaystyle\frac{m_{\tau}c_Lc_R}{v}$ &
$\displaystyle\frac{m_{\tau}c_Lc_R}{v}$ \\
$H \bar{L}       L$ & $\displaystyle\frac{y_{34}}{\sqrt{2}}s_Lc_R$ &
$\displaystyle\frac{y_{43}}{\sqrt{2}}s_Rc_L$ \\
$H \bar{L}    \tau$ &
$\displaystyle\frac{y_{33}}{\sqrt{2}}s_Lc_RP_R+\frac{y_{34}}{\sqrt{2}}c_Lc_RP_L$
&
$\displaystyle\frac{y_{33}}{\sqrt{2}}s_Lc_RP_R+\frac{y_{43}}{\sqrt{2}}c_Lc_RP_R$
\\ \hline
$Z \bar{\tau} \tau$ & $\displaystyle
-\frac{g}{2\cos\theta_W}\gamma_{\mu} (2\sin^2\theta_W-P_Lc_L^2)$ &
$\displaystyle
-\frac{g}{2\cos\theta_W}\gamma_{\mu}\left(2\sin^2\theta_W
-\frac{1}{2}(1+s_R^2)+\frac{1}{2}c_R^2\gamma_5\right)$ \\
$Z \bar{L}       L$ & $\displaystyle
-\frac{g}{2\cos\theta_W}\gamma_{\mu} (2\sin^2\theta_W-P_Ls_L^2)$ &
$\displaystyle -\frac{g}{2\cos_W}\gamma_{\mu}\left(2\sin^2\theta_W
-\frac{1}{2}(1+c_R^2)+\frac{1}{2}s_R^2\gamma_5\right)$ \\
$Z \bar{L}    \tau$ & $\displaystyle
\frac{g}{2\cos\theta_W}\gamma_{\mu}(P_Lc_Ls_L )$ &
$\displaystyle -\frac{g}{2\cos\theta_W}\gamma_{\mu}(P_Rc_Rs_R)$ \\
\hline \hline
\end{tabular}
\caption{Higgs-fermion-fermion and Z-fermion-fermion interactions in
the vectorlike singlet model and the vectorlike doublet model.
$P_{L,R}=\displaystyle\frac{1\mp\gamma_5}{2}$, $s=\sin\theta$,
$c=\cos\theta$.}
\end{table}

The main phenomenological constraints come from the branching ratio of
$Z\rightarrow \tau\tau$.  Non-vanishing $\sin\theta_L$ results in that
$Z\rightarrow \tau\tau$ deviates from SM prediction.  The current
experimental data and SM prediction are \cite{PDG}
\begin{equation}
\begin{array}{lll}
\Gamma^{\rm exp}(Z\to\tau\tau)& = &(84.09 \pm 0.2)\, {\rm MeV}\,, \\
\Gamma^{\rm SM} (Z\to\tau\tau)& = &(83.82 \pm 0.1)\, {\rm MeV}\,.
\end{array}
\end{equation}
Considering the central value difference and $3\sigma$ uncertainties
of both experimental and theoretical results, the $Z\to\tau\tau$
decay width still allows its one percent at most coming from new
physics (which corresponds to $\sim 0.3\%$ of the branching ratio).
Requiring the uncertainty of the $Z\rightarrow \tau\tau$ width being
smaller than $1\%$, we get $\sin\theta_L < 0.0686$ in the vectorlike
singlet case.

New interactions $Z\bar{L}\tau$ and $H\bar{L}\tau$ provide the mechanism
for single production of exotic leptons via gluon fusion.  We perform
calculation with LoopTools \cite{LoopTools}.  The results of the cross
section are shown in Fig. 7 by taking $\sin\theta_L = 0.05$.  They
include both $\bar{\tau}L$ and $\tau\bar{L}$ production.  The figure
also shows that of the Z boson mediated Drell-Yan process for comparison.
We see that Drell-Yan always dominates over gluon fusion.  For
$m_L=150-250$ GeV, the cross section via gluon fusion is about 0.3 fb
while that via Drell-Yan is several fb which is marginally within the
detect ability at the LHC.  For $m_L > 250$ GeV, even the Drell-Yan
cross section is smaller than 1 fb, this is small for such heavy lepton
detection.

One way to enhance the gluon fusion mechanism is to consider an
additional generation of sequential fermions with large Yukawa
couplings. Namely the physics is the SM plus a fourth chiral
generation and the vectorlike singlet charged lepton.  New
sequential quark loops with $m_U=400$ GeV and $m_D=m_U-40$ GeV, for
an example, increase gluon fusion contribution in single production
processes.  Then the cross sections of Higgs mediated gluon fusion
are larger than those of Z boson mediated gluon fusion.  And the
gluon fusion mechanism can dominate over the Drell-Yan mechanism, as
shown in Fig. 7 for $m_L > 350$ GeV.  In this case the cross section
can be as large as 0.3 fb.  This again is still challengingly small
for its detection at the LHC.

\begin{figure}[htbp]
 \includegraphics[width=4in]{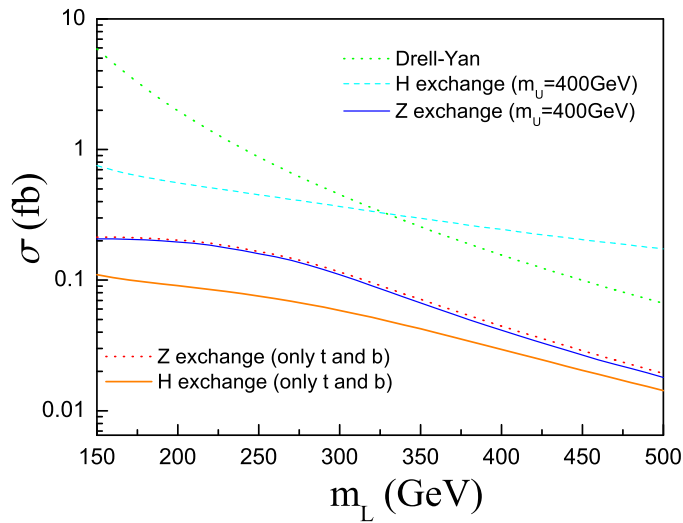}
\caption{Cross sections for vectorlike singlet lepton single
production via the gluon fusion mechanism: Higgs exchange (lower
solid line) and Z exchange (lower dot line) without new sequential
fermions; and Higgs exchange (dash line) and Z exchange (upper solid
line) with additional sequential fermions where $m_U=400$ GeV,
$m_D$=$m_U$-50 GeV. The Drell-Yan mechanism (upper dot line) is
given for comparison.}
\end{figure}

Let us make few remarks.  (1) Compared to heavy sequential lepton
pair production studied in the last section, the vectorlike lepton
single production rate is small, in spite of the phase space
enhancement.  This is mainly due to that we have used full loop
calculation.  The smallness is also due to suppression of
$\sin\theta_L$ which is strongly constrained by the branching ratio
of $Z\rightarrow \tau\tau$.  Note that we have used $\sin\theta_L$
being 0.05 which is just half of that adopted in previous studies
\cite{singlelepton}.  (2) For vectorlike lepton pair production,
because $HLL$ interaction and axial vector current $ZLL$ interaction
are proportional to $\sin\theta_L$ and $\sin\theta_L^2$,
respectively, in this model, the cross sections of the Higgs
exchange diagram and the Z exchange diagram are suppressed
significantly by $\sin\theta_L$ in certain power.  (3) The
phenomenology analysis for vectorlike doublet lepton models is
similar to the singlet case. The production results are similar to
the above singlet scenario. So we will not discuss the doublet
lepton scenario further.

\section{Conclusion}

In this paper we have revisited heavy lepton productions at the LHC.
Our focus is the gluon fusion mechanism which can be important due
to large rate of gluons at the LHC.  If contribution from new
generation quarks is considered, the cross sections via the gluon
fusion mechanism can be enhanced significantly.  The pair production
of new sequential heavy leptons from gluon fusion at the LHC
dominates over that of the Drell-Yan mechanism in the large lepton
mass region.  With a luminosity of 100 fb$^{-1}$, we predict that
for the sequential lepton mass $m_L=250$ GeV, 8100 heavy charged
lepton pair events can be produced at the LHC with $\sqrt s$=14 TeV.

We have also calculated exotic lepton single production in
vectorlike lepton extended models.  In the gluon fusion mechanism,
we have included the Higgs exchange. However, the production rate
for exotic lepton is small due to suppression of the mixing
parameter. Our numerical results for both pair and single production
of heavy leptons are smaller than previous studies especially for
the heavy lepton in the large mass region.  The main reason is that
we have not used tree level approximation.  In the loop computation,
we have also adopted updated parton distribution function and new
electroweak physics data.

\acknowledgments

The authors would like to thank Tao Han, Zong-guo Si, Wen-Long Sang
and Lei Wang for helpful discussions.  This work was supported in
part by the National Science Foundation of China under Grant Nos.
90503002 and 10821504, and by the National Basic Research Program of
China under Grant No. 2010CB833000.

\appendix*

\section{}
The cross section for the 2-2 process at hadron colliders is \beq
\sigma(P_AP_B\to F_3F_4) = \sum_{a,b} \int dx_1 dx_2\
f_{a/A}(x_1,Q^2) f_{b/B}(x_2,Q^2) \frac{P_{out}}{32 \pi^2
\hat{s}^{3/2}}\overline{|\mathcal{M}|}^2 d\Omega\ , \eeq where
$P_{out}=\sqrt{\frac{(\hat{s}+m_4^2-m_3^2)^2}{4\hat{s}}-m_4^2}$, and
$m_3$ and $m_4$ are the masses of final states $F_3$ and $F_4$,
respectively. For the Higgs and Z exchange diagrams of the pair
production and single production of heavy leptons, the Feynman
amplitudes are represented as follows. \bea
&&\mathcal{M}_H=\frac{g_s^2}{4\pi^2v}I_H(g^{\mu\nu}-\frac{2p_1^{\mu}p_2^{\nu}}{\hat{s}})\epsilon_{\mu}(p_1)\epsilon_{\nu}(p_2)\frac{i}{\hat{s}-m_H^2+iM_H\Gamma_H}\frac{m_L}{v}\bar{u}(p_3)v(p_4).
\\
&&\mathcal{M}_Z=F^{\alpha\mu\nu}\epsilon_{\mu}(p_1)\epsilon_{\nu}(p_2)\frac{i(-g_{\alpha\beta}+k_{\alpha}k_{\beta}/{m_Z^2})}
{\hat{s}-m_Z^2+iM_Z\Gamma_Z}\bar{u}(p_3)i\gamma_{\beta}(g_v+g_a\gamma_5)v(p4).
\label{ampz} \eea

 In formula (\ref{ampz}), $F^{\alpha\mu\nu}$ is the ggZ interaction vertex as represented in
formula (\ref{ggZ}). The $I_H$ represented in Passarino-Veltman is:
\beq I_H=\sum_Q m_Q^2 \left[(1+(2m_Q^2-\frac{\hat{s}}{2})\
C0[0,0,\hat{s},m_Q^2,m_Q^2,m_Q^2]\right],\eeq and $F_i's$ in
$F^{\alpha\mu\nu}$ represented in scalar loop functions are: \bea
F_1 &=& -\frac{1}{\hat{s}}(B0[0,m_Q^2,m_Q^2]-B0[\hat{s},m_Q^2,m_Q^2]+1+2C0[0,0,\hat{s},m_Q^2,m_Q^2,m_Q^2]m_Q^2),\\
\nonumber
-F_2 &=& F_3=\frac{2}{\hat{s}}\left[
-\frac{1}{2}(B0[0,m_Q^2,m_Q^2]-B0[\hat{s},m_Q^2,m_Q^2]+1
-2C0[0,0,\hat{s},m_Q^2,m_Q^2,m_Q^2]m_Q^2)+1 \right],
\\
\\
F_4 &=&
-\frac{1}{2}(B0[0,m_Q^2,m_Q^2]-B0[\hat{s},m_Q^2,m_Q^2]+1-2C0[0,0,\hat{s},m_Q^2,m_Q^2,m_Q^2]m_Q^2)+1,
\eea and LoopTools \cite{LoopTools} is used for the numerical
calculation of the scalar loop functions.

The general representations of Passarino-Veltman scalar loop
functions $B0$ and $C0$ are \cite{Sfun}: \bea
&&B0[p_1^2,m_1^2,m_2^2]=\frac{1}{i\pi^2}\int
 d^Dq\frac{1}{[q^2-m_1^2][(q+p_1)^2-m_2^2]}.\\\nonumber
&&C0[p_1^2,p_2^2,(p_1+p_2)^2,m_1^2,m_2^2,m_3]=\frac{1}{i\pi^2}\int
 d^Dq\frac{1}{[q^2-m_1^2][(q+p_1)^2-m_2^2][(q+p_1+p_2)^2-m_3^2]}.\\
  \eea


\end{document}